\documentclass[pre,twoside,twocolumn,floatfix]{revtex4}
\usepackage{amssymb}
\usepackage{amsmath}
\usepackage{amsfonts}
\usepackage{graphicx}
\usepackage{physics}
\usepackage{xcolor}
\usepackage{float}
\usepackage{tikz}
\usepackage{oplotsymbl}
\usetikzlibrary{shapes}
\usepackage{xcolor}
\usepackage[pass]{geometry}
\usepackage[colorlinks=true,linkcolor=blue]{hyperref}
\begin{document}

\title{Electronic screening of the friction acting on ions and water molecules in narrow carbon nanotubes}
\author{A.W.C.\ Lau$^1$}
\author{J.B.\ Sokoloff$^{1,2}$}
\affiliation{(1) Department of Physics, Florida Atlantic University, 777 Glades Road, Boca Raton, Florida 33431}
\affiliation{(2) Physics Department, Northeastern University, Boston, Massachusetts 02115}

\date{\today}

\begin{abstract}
Li {\it et al.} have observed a larger  flow rate, resulting from osmotic pressure, of protons and water molecules in nanometer
scale diameter metallic carbon nanotubes compared to that in semiconducting carbon nanotubes. The flow 
rate of potassium ions, however, under an applied electric field is almost the same in metallic and semiconducting nanotubes.  
We propose a simple physical picture to understand these experimental results by examining the effects of screening by conduction electrons in electrically conducting carbon nanotubes on the friction experienced by protons, water molecules, and ions flowing through the nanotube.
\end{abstract}

\maketitle
%%%%%%%%%%%%%%%%%%%%%%%%%%%%%%%%%%%%%%%%%%%%

\section{Introduction}

Understanding the physics at the nanoscale for device applications, such as filtration devices \cite{hummer,noy2,corry}, fuel cells \cite{schuster,hummer2,hummer3}, and sensing devices \cite{dellago1} requires an understanding of the flow of ions, protons and water molecules through confined narrow tube-like structures. An important aspect is to elucidate the role of electronic degree of freedom in determining the transport properties of carbon nanotubes (CNTs) \cite{quantum, cui}. Whereas these references deal with ions and water in nanotubes with radii greater than a nanometer, our treatment deals with nanotubes with diameters less than $1.3\,\mbox{nm}$.  Recently, Li {\em et al.} \cite{Noy-Nature-Mat} have experimentally studied the effects of the electron conductivity of a CNT on the flow of ions and water through a subnanometer diameter nanotube. They have found that the mobility of water and hydronium and hydroxide ions is larger in metallic nanotubes than its value in semiconducting nanotubes, but the mobility of ions is about the same. In order to interpret their experimental results, molecular dynamics simulations have been performed in which the effects of polarization of the nanotube due to conduction electrons of the tube are treated using the Thole tensor \cite{thole}. It takes into account for the interaction between the dipole moments induced in the carbon atoms due to the electric field of the water and the ions in the nanotube. The polarization of a metallic nanotube is treated by choosing the parameters in the Thole tensor to produce a large value of the component of the polarizability  of the nanotube parallel to its axis. 

In this paper, we discuss the experimental results of Ref.\ \cite{Noy-Nature-Mat} and related issues, in the context of the screening of the interaction of the water molecules and ions in the nanotube with the carbon atoms in electrically conducting nanotubes.  We treat the screening within the Thomas-Fermi approximation, which is valid when electron density is sufficiently high\cite{ashcroft,TF-sim}. Since the band gaps in semiconducting carbon nanotubes are relatively small, it should give qualitatively correct results for semiconducting nanotubes as well. In Section \ref{model}, the experimental methods used to study the flow of ions, protons and water molecules through carbon nanotubes are summarized and an explanation is proposed for why the mobility of ions is almost the same in metallic and semiconducting nanaotubes. In Section \ref{section3}, we argue that the screening due to conduction electrons in semiconducting and in metallic CNTs can account for the larger mobility of protons and water molecules in metallic CNTs than in semiconducting CNTs. Section \ref{section4} discusses the contribution to the friction acting on protons due to electronic excitations in the nanotube.  In Section \ref{section5}, it is shown that the interaction between a proton or ion with nanotube due to the polarization of the carbon atoms by the electric field of the proton or ion is able to lower their energy sufficiently to allow them to reside in the tube. Section \ref{section6}  discusses what we can learn from the existence of proton conductivity in nanotubes containing water wires. In Section \ref{section7}, it is shown that it is possible for charged defects in a water wire, such as hydronium or hydroxide ions to be prevented from flowing with a water wire that is pushed through a nanotube by an applied pressure or osmotic pressure.

\section{Summary of the experimental methods used to measure ion, proton, and water flow through metal and semiconductor nanotubes} 
\label{model}

Let us examine the experimental situation by which the flow of ions, of protons, and of water is measured in Ref.\ \cite{Noy-Nature-Mat}. The  conduction from hydronium  ions is determined by surrounding a pH sensitive material by a lipid membrane penetrated by several nanotubes. Hydronium ions are added
to the region outside of this membrane by adding acid. The number of these ions that have flowed through the nanotubes is determined by measuring the pH inside the lipid membrane. See Fig.\ 3a in Ref.\ \cite{Noy-Nature-Mat}.  

The method of measuring the electrical conduction due to ions is illustrated in Fig.\ 1c in Ref.\ \cite{Noy-Nature-Mat}. Two lipid enclosed droplets are connected by a nanotube. An electrical potential between the droplets is created by placing an electrode connected to a battery in each of the droplets. If the nanotube is an electrical conductor, the region inside the tube is an equipotential region. In the Appendix \ref{appen}, we estimate that the time required for the semiconducting nanotube to become an equipotential region once an electric field is applied is only of the order of $10^{-5}\,\mbox{s}$. Therefore, it is reasonable to assume that the nanotube is an equipotential region during the experiment. Then, the electric field inside the tube is zero for semiconducting in addition to metallic nanotubes. If the applied electric field points from left to right, the left side is negatively charged and the right side is positively charged. Therefore, ions on the left side will be pulled towards the opening of the tube, enter the tube and then drift
through the inside of the tube, where they are subject to zero electric field.

Friction might initially stop the ion in the tube, but it can still diffuse (in both directions) and it can be pushed along if another ion enters the tube and collides with it. 
%Since the gaps in the semiconducting nanotubes are small compared to $k_B T$ and the
%experiments reported in Ref. \cite{Noy-Nature-Mat} are done at room temperature, 
Therefore, it is resonable that the ionic conduction is the same for both metallic and semiconducting nanotubes, because as long as the electronic conductivity of the
nanotube is nonzero, the tube is an equipotential, and hence, the ion is only acted on by the electric field when the ion is outside the nanotube. A possible model for the motion of ions through the nanotube is presented in Appendix B.

The mechanism proposed in Ref.\ \cite{Noy-Nature-Mat} to study the flow of protons through the nanotubes is based on the diffusion
of protons from a region of high concentration (i.e., low pH) to a region of low concentration (i.e.\ high 
pH). Since what is reported in Ref.\ \cite{Noy-Nature-Mat} is diffusion of protons, the measured mobilities must
actually be the mobilities calculated from the Nearnst-Einstein equation. On the other hand, the water flow rate reported in Ref.\ \cite{Noy-Nature-Mat} is measured by osmotic pressure created by placing sugar molecules in the solution outside of the membrane, which pulls water
molecules out of a region surrounded by a lipid membrane through nanotubes that are inserted in the
membrane. Note that sugar molecules
were chosen in the experiment because they are too big to pass through the nanotubes.

Protons and/or water molecules moving through both metallic and nonmetallic carbon nanotube are subject to a force of friction due to the excitation of phonons in the nanotube \cite{Ls}.  It is the result of the interaction between protons and/or water molecules and the electric dipoles induced in the carbon atoms by the electric field of the protons and/or water molecules. In metallic nanotubes, there will also be a contribution to the friction resulting from excitation of the conduction band electrons. We will show later, however, that this contribution to the friction is likely much smaller, and therefore, let us ignore it. Therefore, for protons and water molecules the friction due to phonon excitations should be smaller in metallic
nanotubes because the electric field due to the protons and water molecules is screened.

Reference \cite{yang} provides evidence that the friction acting on water flowing through carbon nanotubes is dominated by forces originating from the ends of the nanotube. 
%This conclusion,  however, is based on MD simulations for nanotubes of diameter greater than a nanometer and Poisson-Boltzmann equation solutions, %for which the water does not form water wires. Our treatment of the flow of water through sub-nanometer diameter carbon nanotubes, however, is %based on the model that assumes that water in a sub-nanometer tube flows as a water wire\cite{hummer2,hummer3,dellago1}. Since viscous shearing as %occurs in water flowing in wider nanotubes, does not occur for water wires, the discussion presented in Ref.\ \cite{yang} is not appropriate for %the study of protons and water molecules  flowing through carbon nanotubes. 
Since the experimentally observed difference between the observed friction acting on water and protons in metallic and semiconducting nanotubes, however, is only a small percentage of the observed friction, it is possible that most of the friction does come from end effects but the difference between the flow rate of ions in metallic and semiconducting nanotubes could still be observable. Whereas the friction from the tube wall depends on the nanotube conductivity, it is not obvious that the entrance and exit resistance does. It is more likely related to the forces experienced by the water as it is forced to flow through the smaller volume of the nanotube. Therefore, any dependence of the flow rate through the nanotube on the nanotube conductivity must result from the force of friction between the water or protons flowing through the tube and the nanotube wall.

\section{Thomas-Fermi screening of 2D electron gas confined on a plane and on the surface of a cylinder}
\label{section3}

In Ref.\ \cite{Noy-Nature-Mat}, screening of the ion's potential due to conduction electrons in the wall of the nanotube is accounted for in the simulations reported in that publication by assuming that the nanotube wall provides a large polarizability along the nanotube $\alpha_{zz}$. This is motivated by Ref.\ \cite{brothers} which shows that the polarizatility of a semiconducting nanotube is inversely proportional to the gap in the nanotube electron energy bands. For a metallic nanotube, the band gap is zero. The polarizability of the nanotube is calculated using the Thole tensor \cite{thole} which was used to calculate  the contribution to the polarizability of the carbon atoms in the nanotube from the interaction between the dipoles induced in the carbon atoms by the potential due to the ion. The Thole tensor contains a correction to the dipole interactions with a parameter that turns them off when it is zero and allows them to have their maximum value when it is infinite. The value of both this parameter and the parameter that gives the magnitude of the interaction between the dipoles induced in the carbon atoms are chosen so as to give a large value for $\alpha_{zz}$ for metallic nanotubes.  Here, we propose a more physically transparent, albeit qualitative, approach to the problem by examining the screening of the interaction between an proton or a water molecule in the nanotube and the carbon atoms of the nanotube within the Thomas-Fermi approximation. 

Let us first consider Thomas-Fermi screening of the electric potential of an ion or proton sufficiently close to the nanotube
wall to approximate the wall by a plane. We will use the treatment of Thomas-Fermi
screening in chapter 17 of Ref. \cite{ashcroft} to treat screening resulting from 
conduction electrons in the wall of the nanotube. 
We write Poisson's equation for the potential $\phi({\bf x})$ at position ${\bf x}$ as\begin{equation}
- \nabla^2 \phi({\bf x}) = 4 \pi \rho_{\mbox{\tiny ind}}({\bf x}) + 4 \pi Ze_0 \delta({\bf r}) \delta (z-h),
\label{1}
\end{equation}
due to a point charge $Ze_0$ located at ${\bf x} = ({\bf 0}, h)$, where $h$ is the distance of the charge from the plane at $z=0$, and $\bf{r}=(x,y,0)$. 
 The quantity $\rho_{\mbox{\tiny ind}}({\bf x})$ is the induced charge density in the 2d electron gas confined to the plane. It can be written as\begin{equation}
 \rho_{\mbox{\tiny ind}}({\bf x}) = -e_0 \left [ n_0(\mu + e \phi({\bf x})  ) - n_0(\mu ) \right ] \delta(z), 
\end{equation}
where $e_0$ is the electron charge and $n_0(\mu ) $ is the electron number density found from the Fermi-Dirac distribution\begin{equation}
n_0(\mu )  = \int \frac{d^2 \bf k }{(2 \pi)^2} \, \frac{1}{e^{\beta [\varepsilon(k) - \mu]} +1 }.
\end{equation}
Expanding $\rho_{\mbox{\tiny ind}}({\bf x})$ to first order in $\phi({\bf x}) $, we find\begin{equation}
\rho_{\mbox{\tiny ind}}({\bf x}) \approx -e_0^2 \left ( \frac{\partial n_0 }{\partial \mu }\right ) \phi({\bf x}) \delta(z),
\end{equation}
Substitute this into Eq.\ (\ref{1}), we find\begin{equation}
- \nabla^2 \phi({\bf x})  + \frac{1}{\lambda_{\mbox{\tiny TF}}}\delta(z) \phi({\bf x}) = 4 \pi Ze_0 \delta({\bf r}) \delta (z-h),
\label{phi-eq}
\end{equation}
where the inverse screening length is given by $ \lambda_{\mbox{\tiny TF}}^{-1} = e_0^2 \left ( {\partial n_0 /\partial \mu }\right ).$
Let us write the solution in terms of its partial Fourier transform
\begin{equation}
\phi({\bf x})  = \int \frac{d^2 {\bf q} }{(2 \pi)^2} \, e^{- i  {\bf q} \cdot {\bf r} }\phi(q, z), 
\end{equation}
 Then $\phi(q, z)$ satisfies\begin{equation}
\left ( - \frac{\partial^2}{\partial z^2 } + q^2 \right )   \phi(q,z)  + \frac{1}{\lambda_{\mbox{\tiny TF}}}\delta(z) \phi(q,z) = 4 \pi Ze_0 \delta (z-h).
\end{equation}
The homogeneous solutions are $e^{-q z}$ and $e^{+qz}$, subject to boundary conditions: \begin{eqnarray}
\left. \partial_z \phi(q ,z) \right |_+ - \left. \partial_z \phi(q ,z) \right |_- &=& \phi(q,0) /\lambda_{\mbox{\tiny TF}}, \nonumber \\
\left. \phi(q ,z) \right |_{z=0^+} &=& \left. \phi(q ,z) \right |_{z=0^-}, \nonumber\\
\left. \partial_z \phi(q ,z) \right |_{z=h^+} - \left. \partial_z \phi(q ,z) \right |_{z=h^-} &=& - 4 \pi Ze_0, \nonumber \\
\left. \phi(q ,z) \right |_{z=h^+} &=& \left. \phi(q ,z) \right |_{z=h^-}. \nonumber 
\end{eqnarray}
We can write $\phi(q ,z) = A e^{+qz} $ for $z< 0$, $\phi(q ,z) = B e^{+qz} + C e^{- qz}  $, for $z>0$, and 
$\phi(q ,z) = D e^{- qz}  $, for $z>h$. Using the above boundary conditions to determine $A, ..., D$ the solution is found to be \begin{equation}
\phi(q ,z)= \frac{ 4 \pi Ze_0 }{q} \left ( e^{-q |z- h|} - \frac{e^{-q |z| - q h}}{1+ q \lambda_{\mbox{\tiny TF}}}\right ).
\end{equation}
For $z=0$, this expression can be written as \begin{equation}
\phi(q ,z=0) =4\pi Ze_0 \left ( \frac{\lambda_{\mbox{\tiny TF}} e^{ - q h}}{1+ q\lambda_{\mbox{\tiny TF}} }\right ). 
\label{phiz=0}
\end{equation}
The inverse Fourier transform is given by \begin{equation}
\phi (r_{||},z=0)=4\pi Ze_0^2 \int_0^{\infty} qdq\int_0^{2\pi}d\phi \frac{\lambda_{\mbox{\tiny TF}} e^{iqr_{||}cos\phi}e^{-qh}}{1+q\lambda_{\mbox{\tiny TF}}}
\end{equation}
or
\begin{equation}
\phi (r_{||},z=0)=\frac{4\pi Z e_0^2\lambda_{\mbox{\tiny TF}}}{r_{||}}\int_0^{\infty} du\frac{u\,J_0 (u)}{\lambda_{\mbox{\tiny TF}} u+ r_{||}}e^{-uh/r_{||}},
\label{phi-r}
\end{equation}
where $r_{||}=(x^2+y^2)^{1/2}$ and $u=qr_{||}$. It is interesting to note that at large values of $r_{||}$, $\phi (r_{||},z=0)$ scales as $r_{||}^{-3}$, as can be confirmed by numerical integration of Eq.\ \eqref{phi-r}, instead of an exponential decay for the usual screening in 3d.  We see from Eq.\ \eqref{phi-r} that since $\lambda_{\mbox{\tiny TF}}$ is inversely proportional to the density of states of the carbon nanotube at the Fermi level, $\phi (r_{||},z=0)$ will be smaller in metallic than in semiconducting nanotubes. This implies that the mobility of proton and water is higher in metallic than in semiconducting nanotubes. 

Next, we consider the problem of an ion or proton in a cylinder, which is more appropriate for proton and ions inside the nanotube. Let the $z$-axis be the 
axis of the cylinder. Using an analogous argument leading to Eq.\ (\ref{phi-eq}), we obtain\begin{eqnarray}
- \nabla^2 \phi({\bf x})  &+& \frac{1}{\lambda_{\mbox{\tiny TF}}}\delta(r- r_0) \phi({\bf x}) \nonumber \\
&=& 4 \pi Ze_0 \frac{\delta(r-r')}{r'}\delta(\theta-\theta')\delta (z),
\end{eqnarray}
where $r_0$ is the radius of the cylinder. So, $\phi({\bf x}) $ is the electrostatic potential at point ${\bf x}$ due to a point charge at ${\bf x}'= (r', \theta',0)$
in the presence of a 2d electron gas confined to 
the surface of a cylinder.  Let us write\begin{equation}
\phi({\bf x})  = \int \frac{d q}{2 \pi} \, e^{ i  q z} \sum_{m} \frac{1}{2 \pi}\,e^{ i m (\theta -\theta')}\,\phi_m(q, r), 
\label{phi-x}
\end{equation}
and we find $\phi(q, r)$ satisfies\begin{eqnarray}
\left ( - \frac{1}{r}\frac{\partial}{\partial r} r \frac{\partial}{\partial r} + \frac{m^2}{r^2} + q^2 \right )   \phi_m(q, r)\phantom{++++++++} \nonumber \\
+ \frac{1}{\lambda_{\mbox{\tiny TF}}}\delta(r- r_0) 
\phi_m(q, r) = 4 \pi Ze_0 \delta(r-r')/r'.
\end{eqnarray}
The homogeneous solutions are $K_m(q r)$ and $I_m(qr)$, the modified Bessel's functions of order $m$. 
So, we write $ \phi_m(q, r) = A I_m(qr)  $ for $ 0< r< r'$, $ \phi_m(q, r) = B I_m(qr)  + C K_m(qr) $ for $ r'< r< r_0$ and $ \phi_m(q, r) = D K_m(qr)$ for $r > r_0$. The boundary conditions are\begin{eqnarray}
\left. \partial_r \phi_m(q ,r) \right |_{z=r'^-} - \left. \partial_r \phi_m(q ,r) \right |_{r=r'^+} &=& 4\pi Ze_0/r', \nonumber \\
\left. \phi_m(q ,r) \right |_{z=r'} &=& \left. \phi_m(q ,r) \right |_{r=r'}, \nonumber \\
\left. \partial_r \phi_m(q ,r) \right |_{z=r_0^+} - \left. \partial_r \phi_m(q ,r) \right |_{r=r_0^-} &=& \phi_m(q,r_0) /\lambda_{\mbox{\tiny TF}}, \nonumber \\
\left. \phi_m(q ,r) \right |_{z=r_0^+} &=& \left. \phi_m(q ,r) \right |_{r=r_0^-}. \nonumber 
\end{eqnarray}
Applying the boundary conditions, we find\begin{eqnarray}
\phi_m(q ,r) &=&  4 \pi Ze_0 I_m(qr') K_m(qr)  -  \frac{ 4 \pi Ze_0   r_0}{\lambda_{\mbox{\tiny TF}}} \nonumber  \\
&&  \times \,
\frac{ K_m^2(qr_0)I_m(qr')I_m(qr) } {1+ ({r_0/ \lambda_{\mbox{\tiny TF}}})I_m(qr_0)K_m(qr_0)},
\label{phi_m}
\end{eqnarray}
for $r'<r< r_0$. We see from Eqs.\ \eqref{phi-x} and \eqref{phi_m} that  $\phi(z,r_0)\rightarrow 0$  as $\lambda_{\mbox{\tiny TF}}\rightarrow 0$, implying that $\phi(z,r_0)$ is smaller for metallic than for semiconducting nanotubes. 
For an ion or proton at the center of the cylinder, $r' \rightarrow 0$, and hence,  $I_m(qr')$ goes to zero except for $m=0$ where $I_0(0) =1$. So,  $I_m(qr') = \delta_{m,0}$ as $r' \rightarrow 0$.
So, if the point charge is located at the origin, we have\begin{eqnarray}
\phi(q ,r) &=&  4 \pi Ze_0  K_0(qr) \nonumber \\
&-&  \frac{ 4 \pi Ze_0   r_0}{\lambda_{\mbox{\tiny TF}}} 
\frac{ K_0^2(qr_0)I_0(qr) } {1+ ({r_0/ \lambda_{\mbox{\tiny TF}}})I_0(qr_0)K_0(qr_0)}.
\label{phi-cyl}
\end{eqnarray}
The potential as a function of $z$ is given by 
\begin{equation}
\phi (z,r)=\pi^{-1}\int_0^{\infty} dq\, \cos(qz)\,\phi (q,r).
\label{phi-z}
\end{equation}
Interestingly, we find that for large $|z|$, 
\begin{equation}
\phi(z,r_0)\propto 1/|z|^{1.4},
\end{equation}
using numerical integration of Eq.\ \eqref{phi-z}.

Using the expression for the screening length $\lambda_{\mbox{\tiny TF}}$ given under Eq.\ \eqref{phi-eq} and the fact that the electron energy as a function of its wave vector $k$ near the  Fermi energy where the two branches of the conduction band, given by $\pm \hbar v_F k$, meet, where $v_F$ is the Fermi velocity \cite{charlier}, we find that for a metallic carbon nanotube, $\lambda_{\mbox{\tiny TF}}$ is given by 

\begin{eqnarray}
\lambda_{\mbox{\tiny TF}}^{-1}&=&e_0^2\frac{\partial}{\partial\mu}\int \frac{d^2 {\bf k}}{(2\pi)^2}\, \left[ \frac{1}{e^{\beta (\hbar v_F k-\mu)}+1} \right. \nonumber \\
&&\hphantom{+++++++++}\left. + \frac{1}{e^{\beta (-\hbar v_F k-\mu)}+1} \right ] \nonumber \\
&=&-{e_0^2}\int_0^{k_{\mbox{\tiny max}}} \frac{kdk}{(2\pi)}\frac{\partial}{\partial k} \, \left  [\frac{1}{e^{\beta (\hbar v_F k-\mu)}+1} \right. \nonumber \\
&&\hphantom{+++++++++} \left. -\frac{1}{e^{\beta (-\hbar v_F k-\mu)}+1} \right] \nonumber\\
&=&\frac{e_0^2 k_B T}{\pi (\hbar v_F)^2}\ln(1+e^{\beta\mu}),
\label{lambda}
\end{eqnarray}
where $\beta=1/(k_B T)$, by integrating by parts. Then using $v_F=8\times 10^{5}\,\mbox{m/s}$ \cite{charlier}, we obtain $\lambda_{\mbox{\tiny TF}}=50\,\mbox{\AA}$ for $\mu=0.16eV$. 

For a semiconducting nanotube, the electron energy is given by $\pm \epsilon (k)$, where $\epsilon(k)=[(\hbar v_F k)^2+g^2]^{1/2}$, and hence, $\lambda_{\mbox{\tiny TF}}$ is given by 
\begin{eqnarray}
\lambda_{\mbox{\tiny TF}}^{-1}
&=&-\frac{e_0^2}{2\pi (\hbar v_F)^2} \int_g^{\epsilon_{max}} \frac{\epsilon d\epsilon}{(2\pi)}\,
\frac{\partial}{\partial\epsilon} \, 
\left [ \frac{1}{e^{\beta (\epsilon-\mu)}+1} \right. \nonumber \\
&& \hphantom{++++}\left. -\frac{1}
{e^{\beta (-\epsilon-\mu)}+1} \right ] \nonumber \\
&=& \frac{e_0^2}{2\pi (\hbar v_F)^2} 
\left [k_B T \ln|(e^{-\beta (g-\mu)}+1)(e^{\beta (g+\mu)}+1)| \vphantom{\frac{e^(1)}{1}}\right. \nonumber \\
&& \hphantom{++++}\left. -g\frac{e^{\beta (g-\mu)}-e^{-\beta (g+\mu)}}{(e^{\beta (g-\mu)}+1)(e^{-\beta (g+\mu)}+1)} \right ],
\label{lambda}
\end{eqnarray}
by integrating by parts. Since $e^{\beta g} \gg 1$, for $\mu=0$, we have\begin{equation}
\lambda_{\mbox{\tiny TF}} \approx 
\frac{\pi (\hbar v_F)^2} {e_0^2 g} e^{\beta g}.
\end{equation}

Since the interaction between a proton or a water molecule is reduced by screening due to conduction electrons in the nanotube, we expect that the friction calculated using this interaction will be smaller in metallic nanotubes, which contain more conduction electrons, than in semiconducting nanotubes. The contribution to the friction acting on a proton, for example,  moving through a carbon nanotube, due to the excitation of phonons in the nanotube, as was described in Ref.\ \cite{our-PRL}, results from the  interaction between the proton and the dipole moment induced in each carbon atom by the electric field of the proton. This induced dipole moment is equal to the product of the polarizability of a carbon atom and the electric field at the carbon atom due to the proton. This electric field is just the gradient of the screened potential due to the proton that we calculated above. Therefore, in order to determine the effect of screening on the induced dipole moment, let us calculate the screened electric potential for both a metallic and a semiconducting nanotube from Eqs.\ \eqref{phi-r} and \eqref{phi-z}. 

The electric potential  is shown in Fig.\ 1 for $\lambda_{\mbox{\tiny TF}}/h=10$, which corresponds to the value of $\lambda_{\mbox{\tiny TF}}\approx 50\,\mbox{\AA}$ calculated above and for $\lambda_{\mbox{\tiny TF}}/h=1000$, which corresponds to virtually no screening.   Equation \eqref{phi-r} gives the approximate potential for an ion which is relatively close to the tube wall. The electric potential at the nanotube wall due to an ion at the center of the nanotube calculated from Eq.\ \eqref{phi-z} is shown in Fig. 2 for $\lambda_{\mbox{\tiny TF}}/r_0=10$ and $\lambda_{\mbox{\tiny TF}}/r_0=1000$.  

\begin{figure}[t]
\centerline{\includegraphics[width=0.9 \linewidth]{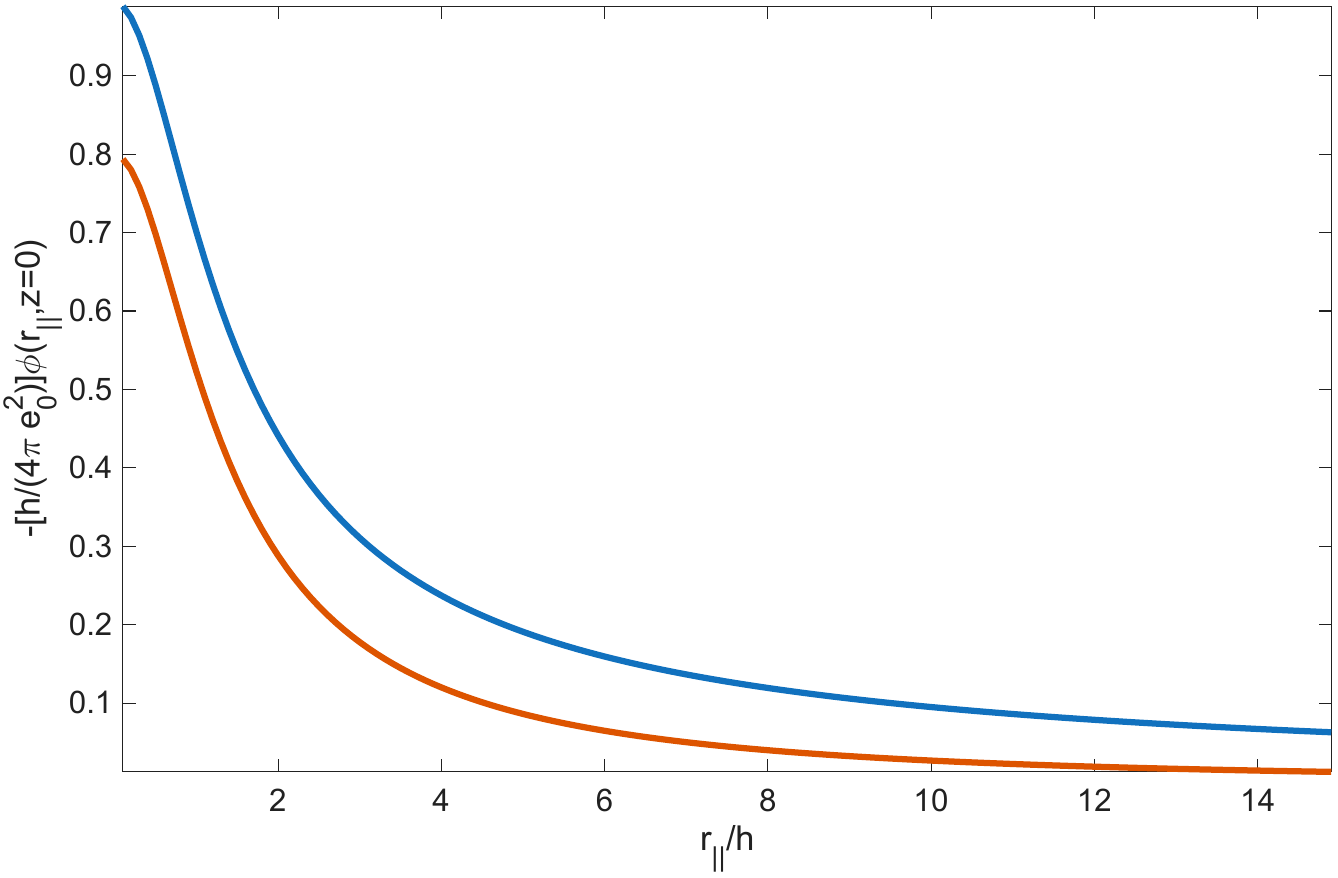}}
\caption{The blue (the upper) figure is a plot of $-\phi (r_{||},z=0)$  versus $r_{||}/h$ for $\lambda_{\mbox{\tiny TF}}/h=1000$ and the red (the lower) figure is a plot of $-\phi (r_{||},z=0)$  versus $r_{||}/h$ for $\lambda_{\mbox{\tiny TF}}/h=10$.}
\label{lambda-1}
\end{figure}

%\begin{figure}[t]
%\centerline{\includegraphics[width=0.9 \linewidth]{plot-close-k_0-0_001.pdf}}
%\caption{A plot of $-d\phi (r_{||},z=0)/dh$  versus $r_{||}/h$ for $\lambda_{TF}/h=1000$. }
%\label{lambda-1000h}
%\end{figure}

%\begin{figure}[t]
%\centerline{\includegraphics[width=0.90\linewidth]
%{plot-close-k_0-0_1.pdf}}
%\centerline{\includegraphics[width=0.90\linewidth]
%{plot-close-k_0-0_001.pdf}}
%\caption{a. A plot of $-d\phi (r_{||},z=0)/dh$  versus $r_{||}/h$ for $\lambda_{TF}/h=10$. b.  A plot of $-d\phi (r_{||},z=0)/dh$  versus $r_{||}/h$ for %$\lambda_{TF}/h=1000$. }
%\label{water}
%\end{figure}
%The force along the tube axis (i.e. the derivative of Eq. \eqref{phi-z} with respect to $z$) is shown in Fig. for $\lambda_{TF}/r_0=10$, which corresponds %to the value of $\lambda_{TF}=50A^o$ calculated above and for$\lambda_{TF}/r_0=100$, which corresponds virtually no screening, as would occur for a %semiconducting nanotube for which $\lambda_{TF}$ is very large compared to the tube radius.  
\begin{figure}[t]
\centerline{\includegraphics[width=0.9 \linewidth]{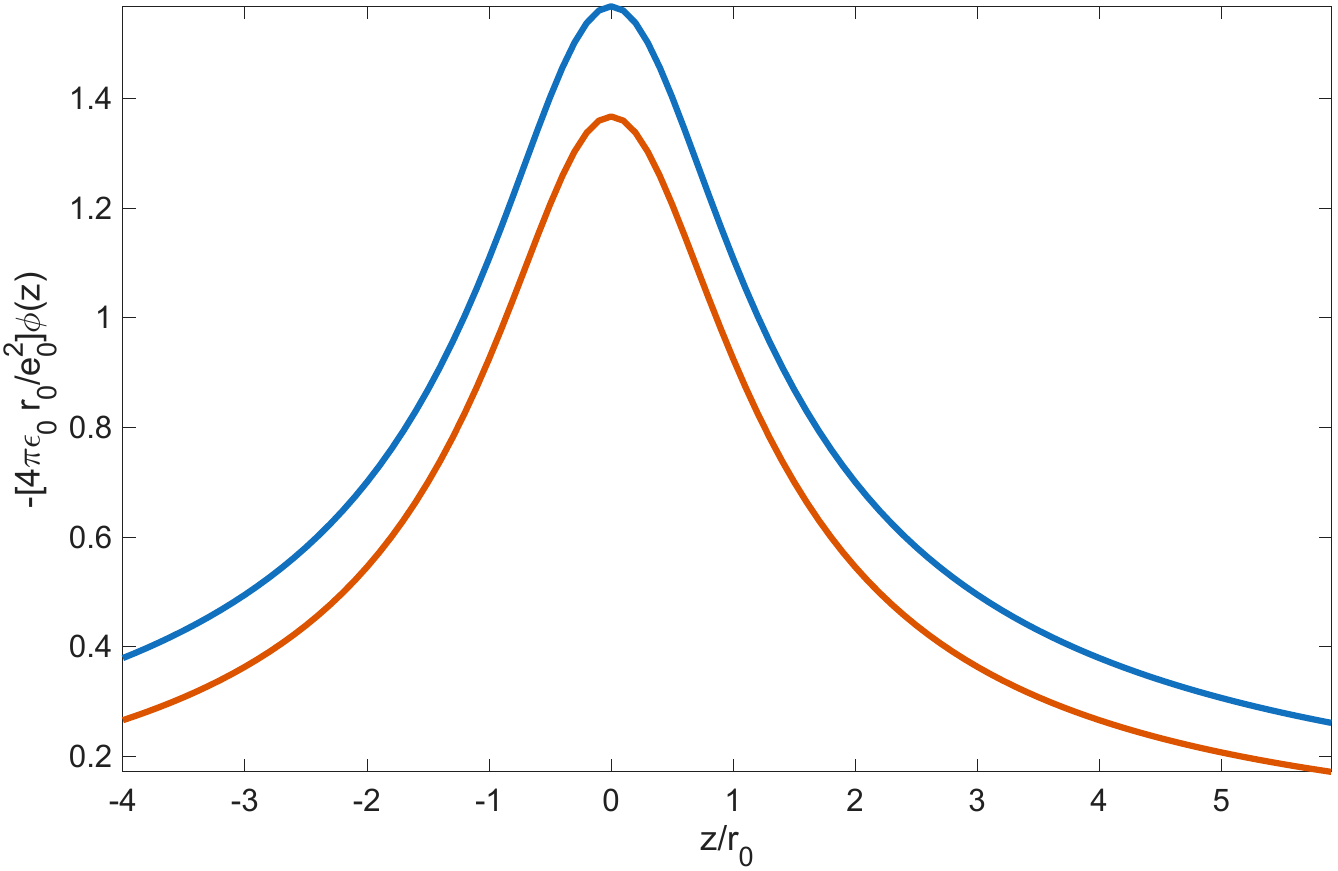}}
\caption{The red (the inner) figure is a plot of $-\phi (z)$  versus $z/r_0$ for $\lambda_{\mbox{\tiny TF}}/r_0=10$ and the blue (the outer) figure is  plot of $-\phi (z)$  versus $z/r_0$ for $\lambda_{\mbox{\tiny TF}}/r_0=1000$.}
\label{lambda-2}
\end{figure}

\begin{figure}[t]
\centerline{\includegraphics[width=0.9 \linewidth]{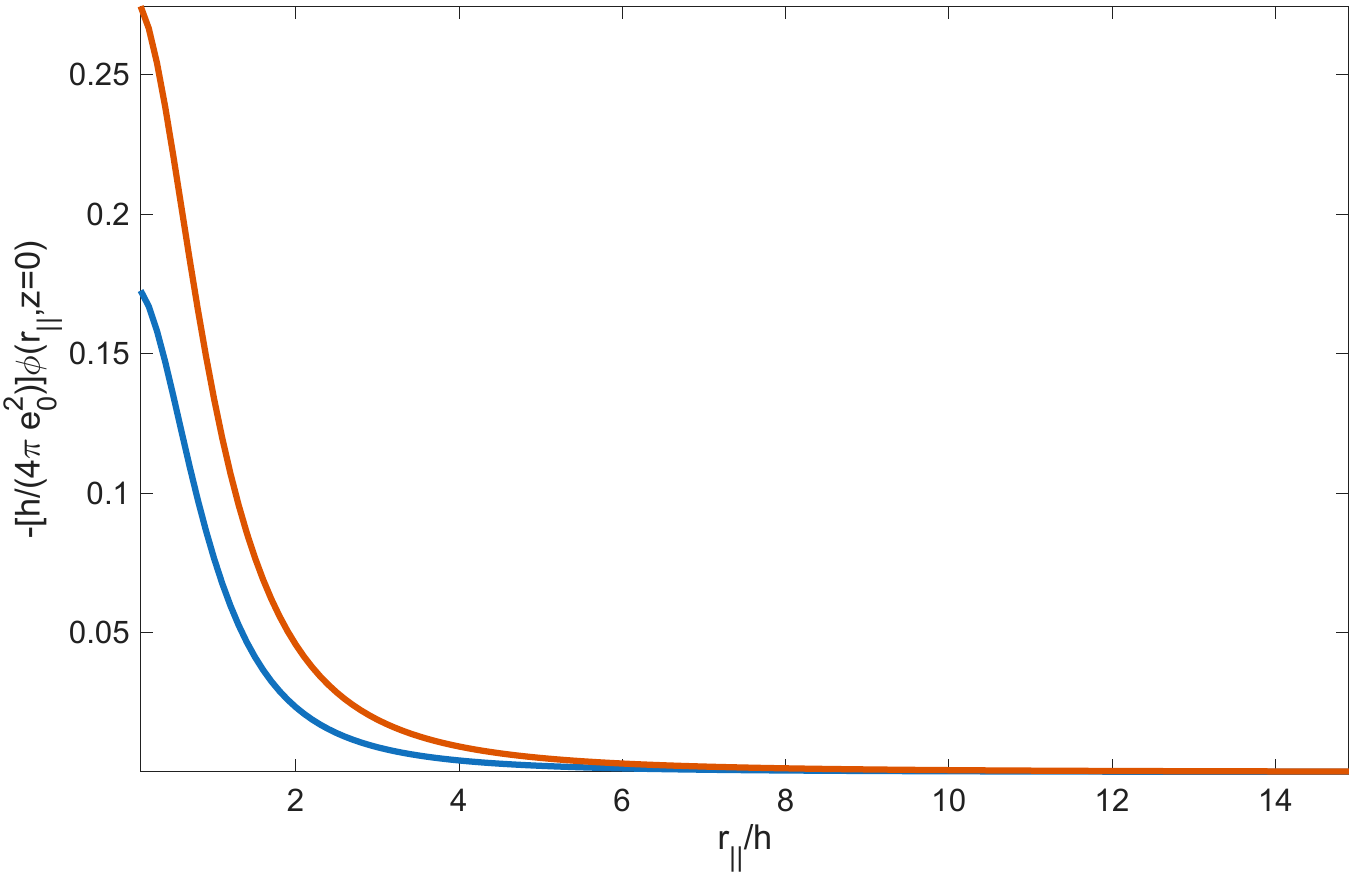}}
\caption{The blue (the lower) figure is a plot of $-\phi (r_{||},z=0)$  versus $r_{||}/h$ for $\lambda_{\mbox{\tiny TF}}/h=0.25$ and the red (the upper) figure is a plot of $-\phi (r_{||},z=0)$  versus $r_{||}/h$ for $\lambda_{\mbox{\tiny TF}}/h=0.5$.}
\label{lambda-3}
\end{figure}

\begin{figure}[t]
\centerline{\includegraphics[width=0.9 \linewidth]{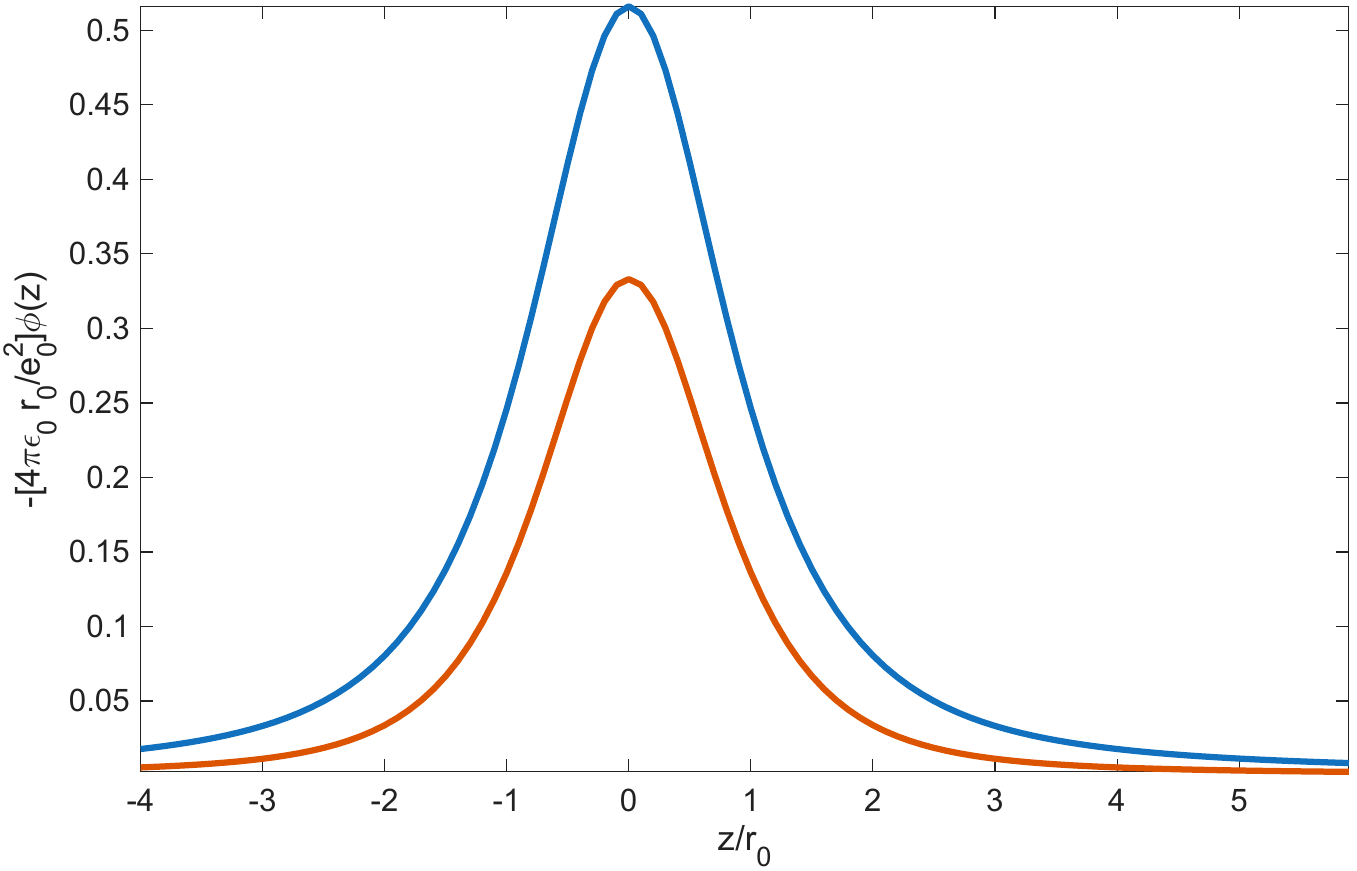}}
\caption{The red (the inner) figure  is a plot of $-\phi (z)$  versus $z/r_0$ for $\lambda_{\mbox{\tiny TF}}/r_0=0.25$ and the blue (the outer) figure is  plot of $-\phi (z)$  versus $z/r_0$ for $\lambda_{\mbox{\tiny TF}}/r_0=0.5$.}
\label{lambda-4}
\end{figure}

We see from Figs.\ \ref{lambda-1} and \ref{lambda-2} that the effect of screening on the electric potential due to the ion is quite small for pure nanotubes, but if the nanotubes in the experiments reported in Ref.\ \cite{Noy-Nature-Mat} were doped with charged impurities that add electrons or holes we will see that the screening has a much larger effect. As an example, let us consider a metallic nanotube doped with charged impurities so that it has enough conduction electrons so that $\lambda_{\mbox{\tiny TF}}=1\mbox{\AA}$. Assuming that $\mu \gg k_B T$, Eq.\ \eqref{lambda} becomes
\begin{equation}
\lambda_{\mbox{\tiny TF}}^{-1}\approx \frac{e_0^2\mu}{\pi (\hbar v_F)^2},
\label{lambda-mu}
\end{equation}
from which we find that $\mu=2.18\times 10^{-19}\,\mbox{J}$. Then, numerically performing the integral in Eq. \eqref{n_0} in the appendix for this value of $\mu$ with $g=0$ we obtain $n_0=3.52\times 10^{18}\,\mbox{m}^{-2}$. For a semiconducting nanotube with $g=0.5\,\mbox{eV}$, we note that since $g \gg k_B T,$ Eq. \eqref{n_0} in the appendix becomes 
\begin{equation}
n_0\approx \frac{\cosh(\beta\mu)}{\pi (\hbar v_F)^2}k_B T ge^{\beta g},
\end{equation}
which gives $\mu=9.88\times 10^{-20} \,\mbox{J}$ for this value of $n_0$. For $\mu>g$ and $\mu-g \gg k_B T,$ Eq. \eqref{lambda} becomes 
\begin{equation}
\lambda_{\mbox{\tiny TF}}\approx \frac{\pi (\hbar v_F)^2}{e_0^2\mu}
\end{equation}
from which we obtain $\lambda_{\mbox{\tiny TF}}=2.21\times 10^{-10}\,\mbox{m}$. Substituting this value of $\lambda_{\mbox{\tiny TF}}$ in Eqs.\ \eqref{phi-r} and \eqref{phi-z}, we obtain the electric potential due to the proton as a function of $r_{||}/h$ for the proton close to the tube wall and $z/h$ for the proton at the center of the nanotube, shown in Figs. \ref{lambda-3} and \ref{lambda-4}.  

Calculations of the screened potential due to a dipole (representing a water molecule) were also done, which give results similar to those shown in Figs. 1-4.

%\section{Ion selectivity issues} 

%In Ref. \cite{tunuguntla}, the
%observation that K+ ions enter a CNT but not the Cl- ions is attributed to the existence of COO-
%ions that form at the ends of the CNT. This point of view is supported by the fact that if you
%reduce the pH of the solution in which the CNT resides (i.e., increase the number of
%protons in the solution), you cause hydrogens in the solution to attach themselves to the
%COO- ions, thus reducing the number of negative ions at the ends of the CNT, the ratio of
%K+ ions to Cl- ions decreases. If one applies a voltage across the CNT to produce ionic
%current through the tube, as illustrated in Fig. 4 of this paper, the barrier due to the 3 COO-
%ions is given by\begin{equation}
%    \frac{3e^2}{4 \pi \epsilon_0 r}= 5.4\,\mbox{eV}
%\end{equation}

%The barrier for Cl- ions calculated by using MD \cite{Noy-Nature-Mat} is about 4.8kcal/mole=0.208eV.
%If 100mV are applied across the CNT, as for example in Fig. 4D of Ref. \cite{Noy-Nature-Mat}, since we saw in section II that most of the charge resides on %the ends of the tube, there will be a
%charge on the end given by\begin{equation}
%    V= EL \approx \frac{Q}{4 \pi \epsilon_0 L}
%\end{equation}
%which gives $Q= 1.1 \times 10^{-21}\,\mbox{C}$, which is two orders of magnitude smaller than the charge due to the $\mbox{COO}$-ions.

\section{Friction due to electronic excitations}
\label{section4}

Let us consider the case of a proton close to the nanotube wall, in which the wall can be approximately treated as a plane. The electronic friction force due to the excitation of conduction electrons in the nanotube by a moving charge is given by \cite{persson,tobin,persson1,sokoloff}
\begin{eqnarray}
F_{\mbox{\tiny fric}}v &=& \frac{2 \pi}{\hbar}\,\sum_{k_i<k_F,k_f>k_F} \left |M_{k_f,k_i} \right |^2 \left [\epsilon(k_f)-\epsilon(k_i) \right ] \nonumber \\
& &\times \delta \left [\epsilon(k_f)-\epsilon(k_i)-\hbar v (k_f-k_i) \right ], 
\end{eqnarray}
treating the electrons in the nanotube as free particles with $v=$ the velocity of the ion or proton, $k_F=(2m\epsilon_F/\hbar^2)^{1/2}$, where $\epsilon_F$ is the Fermi energy, $m$ is the electron mass and the matrix element 
\begin{equation}
 M_{k_f,k_i}=\phi (K,z=0)/A,
 \end{equation}
 where $K=k_f-k_i$, the final minus the initial wave vector, and $A$ is the area of the nanotube.
 The above expression for $F_{\mbox{\tiny fric}}v$ , which is the rate at which the potential due to a proton moving with a velocity v acting on the conduction bands of the nanotube =$\phi 
 (|{\bf r}_{||}-vt {\bf \hat{x}}|,z=0)$ does work on the carbon atoms, where $\hat{x}$ is the  direction of the water flow. We are making the assumption that for a metallic nanotube, the energy bands of a metallic nanotube consist of a conduction band, which we are approximating by free electrons, and some flat bands that can be approximated by individual carbon atom states. 

 From Eq.\ (\ref{phiz=0}), we have
 \begin{equation}
     \phi(K,z=0)=4\pi Ze_0 \left [\frac{\lambda_{\mbox{\tiny TF}}e^{-Kh}}{1+K\lambda_{\mbox{\tiny TF}}} \right ].
     \label{M}
 \end{equation}
For a good metal, $K\approx k_F\approx\pi/a$, $Kh\approx \pi h/a$, and hence, $e^{-Kh} \ll 1$, which suggests that friction due to energy imparted to the conduction electrons in the nanotube by the moving ion could be negligibly small. Since Eq. (\eqref{M}) also becomes small for small values of $\lambda_{\mbox{\tiny TF}}$ for $r=r_0$, it is clear that the electronic friction will also become small for small values of $\lambda_{\mbox{\tiny TF}}$.

 The expansion of Eq.\ \eqref{phi_m} with $r=r_0$ to first order in  $\lambda_{\mbox{\tiny TF}}/r_0$ gives 
\begin{eqnarray}
\phi_m (K,r_0) &\approx& 4\pi Ze_0\left (\frac{\lambda_{\mbox{\tiny TF}}}{r_0} \right )\frac {I_m (Kr')}{I_m (Kr_0)} \nonumber \\
 &\approx& 4\pi Ze_0 \left (\frac{\lambda_{\mbox{\tiny TF}}}{r_0} \right )^{1/2} e^{-K(r_0-r')} \nonumber
 \end{eqnarray}
for $Kr_0 \gg 1$ and $Kr' \gg 1$. Therefore, for sufficiently small $(\lambda_{\mbox{\tiny TF}}/r_0)$ and $e^{-K(r_0-r')}$, the contribution from electronic friction should be small. 

There is also a contribution to the friction due to the dragging of the electrical image produced by the moving ion, which is given in MKS units by \cite{boyer,tomassone,bruch} 
\begin{equation}
F_{\mbox{\tiny fric}}=\frac{e_0^2 v}{16\pi\sigma_{2d} h^2}.
\end{equation}
For $\sigma_{2d}=0.4\times 10^{-4}\,\Omega^{-1}$, $h=2\times 10^{-10}\,\mbox{m}$, and $F_{\mbox{\tiny fric}}/v=3.18 \times 10^{-16}\,\mbox{N}$. This value for the friction is very small compared to the friction due to phonon excitations that we found for an ion moving in a nanotube, $\Delta=10^{-11}\mbox{N}$ \cite{our-PRL}. Note that we have used $\sigma_{2d}=0.4\times 10^{-4}\,\Omega^{-1}$, which is the minimum electrical conductivity, but since the conductivity of a good metal is larger, this contribution for the electronic friction should be smaller. Thus it is reasonable to assume that the contribution to the friction due to interaction of the ion with the conduction electrons will be less important than the contribution from phonon excitations \cite{our-PRL}.  

\section{Ability of ions and protons to reside inside the nanotube}
\label{section5}

The attractive polarization interaction of an ion or proton with the CNT is of the order of\begin{equation}
     |V_i|=\frac{\alpha_{\mbox{\tiny eff}} e_0^2 }{ \left [ r_i^2 +(z-z_i)^2 \right ]^2} \leq \frac{\alpha_{\mbox{\tiny eff}} e_0^2 }{ r_i^4} = 0.0385 \,\mbox{eV},
\end{equation}
where $\alpha_{\mbox{\tiny eff}}$ is the effective polarizability of a carbon atom in the nanotube (determined in Ref.\ \cite{our-PRL}), which is much larger in magnitude than the Lennard-Jones interaction energy parameter ($4.96 \times 10^{-3}\, \mbox{eV}$) used in
Dellago’s simulations \cite{hummer3}. Since in the simulations of Ref.\ \cite{hummer3} both ions and protons are able to
enter the CNT, the polarization interaction is clearly strong enough overcome the increase in energy of an ion or proton due to the loss of water molecules in the hydration shell, allowing ions to enter the nanotube. 

\section{What the observation of conduction due to protons tells us about water in narrow nanotubes}
\label{section6}

Although the simulations reported in Ref.\ \cite {combre} show that at room temperature the water wire is broken up into individual water molecules that are not hydrogen bonded together, as they are in a hydrogen bonded water wire, the observation of diffusion of protons through a carbon nanotube from a region of low $\mbox{pH}$ to a region of higher $\mbox{pH}$ in Refs.\ \cite{Noy-Nature-Mat} and \cite{Tunuguntla} provides evidence that in fact the water wires are hydrogen bonded together and the protons flow through them via the Grotthuss mechanism \cite{grotthus}. Furthermore, the measured rate of diffusion of protons through a $0.81\mbox{nm}$ diameter nanotube, in which it is known that the water flows through it as hydrogen bonded water wires, is much higher than it is in $1.5\mbox{nm}$ diameter nanotubes, in which the water flows as bulk water. This provides additional evidence that the water molecules  in 0.81nm diameter nanotubes flow as  hydrogen bonded water wires \cite{Tunuguntla}.   The proton conductance calculated from the measured diffusion  is of the order of  $10^{-10}\,\mbox{S}$ \cite{Tunuguntla}. 

%Dellago finds that if excess protons can enter a nanotube, their diffusion constant is \begin{equation}
%    D= 0.17\, \mbox{nm}^2/\mbox{ps} = 0.17 \times 10^{-6}\,\mbox{m}^2/\mbox{s}.
%\end{equation}
%Since \begin{equation}
 %   \frac{D}{\mu\,k_B T }= R_{\mbox{\tiny NE}},
%\%end{equation}
%if $R_{\mbox{\tiny NE}}=1$, $\mu = 6.87\, \mbox{m}^2/\mbox{(N$\cdot$s)}$. Then, the conductivity is given by \begin{equation}
 %  \sigma = n_{\mbox{\tiny H}} e \mu \pi R^2 = 3.39 \times 10^{-19}\,\mbox{Am}^2/\mbox{V},
%\end{equation}
%where for a pH of 3, the number density of protons is given by\begin{equation}
 %  n_{\mbox{\tiny H}} = 6.2 \times 10^{23}\,\mbox{m}^{-3},
%\end{equation}
%when the pH=3 on one side. Then, the resistance of a $10$ nm long nanotube is given by\begin{equation}
 %  R= L/\sigma= 2.95 \times 10^{10}\,\mbox{$\Omega$}.
%\end{equation}
%If $100$ mV is applied across the nanotube, the current due to the protons would be given by\begin{equation}
 %  I = V/R= 3.39\, \mbox{pA}.
%\end{equation}
%which is a little larger than the values given in Fig. 4D of Ref. \cite{Noy-Nature-Mat}, but still of the correct order of magnitude. If $R_{\mbox{\tiny %NE}}$ were equal to $1/5000$, $\mu$ would be a factor of 5000 larger, which means that $\sigma$ , and hence, $R$ will be a factor of 5000 smaller, making %$I$ a factor of 5000 larger.

Whereas ions flow through the nanotube, protons under an applied field move by undergoing
successive hops between water molecules in a water wire. This will result in an average dry (velocity independent) friction, as opposed to friction proportional to the flow velocity.

\section{Falling behind of charged defects in water wires}
\label{section7}

The treatment in Ref. \cite{Ls} of the contribution to the force of friction between a water structure and the carbon nanotube in which it is moving from defects in the water structure assumes that these defects are stationary with respect to the water wire. It will be argued in this section that at least some of these defects might not be stationary with respect to a water wire, but instead, might actually be left behind as the water wire moves. For example, the charged defects (both hydronium and hydroxide) in water wires could get left behind when the
water wire is pushed through the nanotube by a pressure gradient.  This is because the mobility of these
defects relative to the water structure or water wire can be large. The mobility of a proton gives the velocity relative to the water wire of the defect induced by an applied force (in this case, the friction  due to its interaction with the nanotube). If the water wire is made to flow through
the nanotube by a pressure gradient, the friction force exerted on the defect by the nanotube could
prevent the defect from moving along with the water. 

Let $\mu_0$ be the mobility due to proton hopping relative to the water wire, which does not include the friction due to the nanotube wall. There is a friction force holding the protons back, $F_1 = F_d /(n N_w)$ \cite{Ls}, where $F_d$ is the force of friction acting on the water wire due to the protons or hydroxide ions moving with it, $N_w$ is the number of water molecules in the water wire and $n$ is the concentration of protons or hydroxide ions. The proton or hydroxide ion will be completely left behind if 
 $F_1\mu_0 > v$. We found that $F_d/nN_w = 3.1 \times  10^{-12}\, \mbox{N}$ \cite{Ls}.  
Reference \cite{dellago3} gets a diffusion constant for the proton and hydroxide defects of $D=  0.17 \times 10^{-6}
\,\mbox{m$^2$/ s}$. Of course, the force of friction $F_d$ should be reduced in metallic nanotubes because of screening by the conduction electrons, reducing the possibility of the protons being left behind because of the friction acting on them due to their electrostatic interaction with the nanotube wall. If we
assume, as is assumed in the Ref.\ \cite{Noy-Nature-Mat}, that the Nernst-Einstein relation holds
for charged defect motion, which should be the case because this mobility does not include friction
due to the wall,\begin{equation}
\mu_0 =D/k_B  T = 4.25 \times 10^{13}\, \mbox{m/Ns}
\end{equation}
and hence, $\mu_0 F_1 = 131 \,\mbox{m/s}$. Therefore, the above inequality holds, implying that the defect is left behind. 

According to Ref.\ \cite{dellago3}, the charged defects can only flow relative to the water wire if the overall polarization of the wire is in the correct direction and flow of these defects reverses the polarization of the chain. The dry friction acting on
the defects from the nanotube pulls both positive and negative defects in the same direction
relative to the water wire. From Fig.\ 4B in Ref.\ \cite{dellago3}, we can see that when the friction
pulls a positive defect to the right, an additional water dipole is turned to the left, but when the
friction pulls a negative defect to the right, an additional defect is turned to the right. Thus, if we
have the same number of positive and negative charged defects in the chain, the net polarization is
maintained, and hence, the defects can continue to move relative to the water wire. Thus, defects
will remain stationary, as the water wire moves. The net friction, due to the defects, acting on the
water wire in this case, is equal to $v/ \mu_0 $ {\em i.e.}, it is viscous friction. If the chain has a net charge,
however, (i.e., there is an excess of defects of one charge), the motion of the defects relative to the
nanotube will eventually polarize the chain in a direction that prevents further motion of the defects
relative to the water wire. At this point the defects will be dragged along with the water wire, and
our calculation of the friction in Ref.\ \cite{Ls} will be correct. Orientational defects have a diffusion constant that is
a factor of three smaller, giving $\mu_0 F_d \approx 45 \,\mbox{m/s}$, which says that they will also not move with a flowing water chain. This behavior will probably only occur for water wires but not for water cylinders, since water cylinders are two dimensional surfaces which have each water molecule
hydrogen bonded with four neighboring molecules.

\section{Conclusion} 
 
We have interpreted the experimental results of Ref.\ \cite{Noy-Nature-Mat} on the enhancement of the mobility of water and hydronium and hydroxide ions in metallic carbon nanotubes on the basis of the electronic screening of the interaction between the  water and hydronium and hydroxide ions with the nanotube. The experiments described in Ref.\ \cite{cui} on the flow of ions in solution in $100\mu m$ long carbon nanotubes with radii of $1.3$\,nm to $2.7$\,nm show, in apparent contrast to the measurements reported in Ref.\ \cite{Noy-Nature-Mat}, that the electrical conductivity of the ions is larger in semiconducting than in metallic nanotubes. We believe that this can be accounted for by the fact that the nanotubes Ref.\ \cite{cui} are $100$\,$\mu$m long, whereas the nanotubes in Ref.\ \cite{Noy-Nature-Mat} are only $11$\,nm long. As a consequence, the time that it takes for the semiconducting nanotubes in the experiments reported in Ref.\ \cite{cui}  to become equipotential regions is much longer than for the shorter nanotubes of Ref.\ \cite{Noy-Nature-Mat}, as illustrated in Appendix A. Therefore, the semiconducting nanotubes are not equipotential regions, and there will be a difference between the ionic conductivity in semiconducting and metallic nanotubes. This will be discussed in more details in a future paper.	

\appendix
\section{The time scale by which the interior of a semiconducting CNT becomes an equipotential}
\label{appen}

It was argued in Sec.\ \ref{model} that when an electric field is applied along an electrical semiconducting carbon nanotube, the inside of the nanotube will be an equipotential region. This will be the case if the time that it takes for electrons to flow along the nanotube, in order to cancel the applied electric field (i.e., in order to make the nanotube an equipotential), is sufficiently short. In this appendix, the time that it takes to accomplish this will be estimated. The concentration of conduction electrons and holes in a semiconducting nanotube with a gap $g$ is given by
\begin{equation}
n_0=\int_0^{k_{\mbox{\tiny max}}}
\frac{kdk}{2\pi} \left [\frac{1}{e^{\beta (\epsilon (k)-\mu )}+1}+\frac{1}{e^{\beta (-\epsilon (k)-\mu )}+1} \right ] 
\end{equation} 
where $\epsilon (k)=[(\hbar v_F k)^2+g^2]^{1/2}$. Then, 
\begin{equation}
n_0=\frac{1}{2\pi (\hbar v_F)^2}\int_{g}^{\epsilon_{\mbox{\tiny max}}}\epsilon d\epsilon  \left [\frac{1}{e^{\beta (\epsilon-\mu )}+1}+\frac{1}{e^{\beta (\epsilon+\mu )}+1} \right ]
\label{n_0}
\end{equation}
\begin{equation}
n_0\approx \frac{1}{\pi (\hbar v_F)^2}(g+k_B T)e^{-\beta g}=2.37\times 10^7m^{-2}
\end{equation}
for $g=0.5\,\mbox{eV}$. Let us now estimate the time that it takes for the nanotube to become an equipotential region, which is the time that it takes for a charge $Q$ to flow from one end of the tube to the other, where $Q$ satisfies
\begin{equation}
\frac{Q}{4\pi\epsilon_0 L^2}<E,
\end{equation}
in MKS units, where $E$ is the applied electric field and $L$ is the tube length. As rough approximation, let us approximate the mobility by the Drude model. Then, the current density is given by
\begin{equation}
j=\frac{e\tau n_0}{m}E,
\end{equation}
where $\tau$ is the scattering time of the conduction electrons and $m$ is the electron effective mass. Then, the time $t$ it takes to establish this charge is given by
\begin{equation}
\frac{e\tau n_0}{m} EC_h t=Q<4\pi\epsilon_0 L^2E,
\end{equation} 
where $C_h$ is the nanotube's circumference. Using for $m^*$, the free electron mass, $L=100\,\mbox{nm}$, and using for $\tau=10^{-14}\,\mbox{s}$ the value for a typical metal \cite{ashcroft}, we get $t=0.655\times 10^{-5}\,\mbox{s}$. Thus, the time $t$ is short enough to conclude that the nanotube in the measurement of the ion flow under an applied electric field in Ref.\ \cite{Noy-Nature-Mat} is an equipotential region. 

If $L$ is increased to $10^{-4}m$, the nanotube length in Ref. \cite{cui}, and $C_h$ is increased by a factor of 2, $t=2\times 10^3 s\approx 1hr$. Thus, the semiconducting nanotube will likely not be an equipotential.

\section{Forces Pushing Ions into the Nanotube}

Both a metallic and a sufficiently short semiconducting nanotube in an electric field is an equipotential region, which means that the ions are not pushed through the tube by the electric field inside the nanotube, but instead get pushed through the tube by successive ions that are pulled into the tube. In contrast, the water and protons are pushed through the tube by osmotic pressure. If only one ion can enter the tube at a time, the ion motion through the tube is described by a “billiard ball model,” in which when an ion enters the tube, its speed on entry is rapidly slowed to zero by friction. The next ion that enters the tube imparts its kinetic energy to the ion already in the tube through their screened Coulomb interaction, assuming that the new ion enters the tube after the first ion has stopped or slowed down. If the ion already in the nanotube is initially moving with a velocity $v_{10}$ and the new ion is initially moving with a velocity $v_{20}$ and $v_{20}>v_{10}$, let us account for the friction using the following model: Using conservation of energy and momentum,
\begin{equation}
 v_1^2+v_2^2=K(v_{10}^2+v_{20}^2)   
\end{equation}
\begin{equation}
v_1+v_2=v_{10}+v_{20}
\end{equation}
where $v_1$ is the velocity of the first ion that enters the tube and $v_2$  is the velocity of the second ion after their collision.  Then, substituting for $v_2$ in Eq. (B1) using Eq. (B2), we obtain
\begin{equation}
v_1^2-(v_{10}+v_{20})v_1+v_{10} v_{20}+(1-K)(v_{10}^2+v_{20}^2)=0
\end{equation}
whose solution is
\begin{equation}
 v_1=\frac{v_{10}+v_{20}}{2}\pm [(\frac{v_{10}-v_{20}}{2})^2-(1/2)(1-K)(v_{10}^2+v_{20}^2)]^{1/2}   
\end{equation}
where $0<K<1$ represents the fraction of the initial kinetic energy that remains after the collision. E.g., if $v_{10}=0$, we get
\begin{equation}
v_1=(1/2)v_{20}+[(1/4)v_{20}^2-(1/2)(1-K)v_{20}^2]^{1/2}
\end{equation}
Thus, $v_1<v_{20}$ for $K<1$ and for $K<1/2$, $v_1=(1/2)v_{20}$. For smaller values of K, there is no solution.   For the screened interaction with the ratio of the nanotube radius to the screening length $r_0/\lambda_{TF}=5$, and $r=0.5r_0$, the screened electrical potential of the ion
\begin{equation}
 \phi(z,r_0)\propto 1/z^2  
\end{equation}
The $1/z^2$ dependence of the potential is not surprising for $r_0/\lambda_{TF}=5$, because we have a dipole in this case, formed by the ion and its screening charge in the tube wall. 

If the two ions come to a stop before the next one enters the nanotube, both in a metallic and in a semiconducting nanotube, the flow rate will depend only on the rate at which new ions enter the nanotube. Since the rate at which ions enter the nanotube depends only on electric field and the charge at the end of the tube, it will be the same for metallic and semiconducting nanotubes, because the charge distribution on the walls of the tube is independent of the conductivity of the tube, as long as it is sufficiently large for the nanotube to be an equipotential region throughout the time that the ions are flowing through the tube. Its charge depends only on the applied electric field. The pressure due to the applied electric field acting on the gas of K+ ions in the reservoirs is given by 
\begin{equation}
P=neEh\approx 9.63\times 10^4 Pa
\end{equation}
assuming that the distance between the electrodes $h=10^{-8}m$, $E=10^7 V/m$  and $n=10mM=6.02\times 10^{24}m^{-3}$  is the number density of ions in the reservoir.  Then the force on an ion at the mouth of the nanotube is approximately equal to 
\begin{equation}
P\pi r_0^2=4.84\times 10^{-14}N
\end{equation}
for a nanotube radius of $r_0=0.41nm$. In comparison, the force on an ion near the end of the nanotube due to the electric field is 
\begin{equation}
\approx eE=(1.6\times 10^{-19})(10^7 V/m)=1.6\times 10^{-12}N
\end{equation}
which is a factor of $33$ larger than the force due to the electric field pressure.  This is the force due to the electric field just outside the tube on a potassium ion that happens to be there. The ion as it enters the nanotube is also acted on by the negative charge near that end of the tube. 
%\section{Acknowledgment}

%We thank 


\begin{thebibliography}{0}

%\bibitem{noy}
%Zhongwu Li {\em et al.}, Nature Nanotechnology, 
%{\bf 18}, 177 (2022).

%\bibitem{jeff}
%J.B.\ Sokoloff and A.W.C.\ Lau, Phys.\ Rev.\ E (in press).

%\bibitem{degennes}
%P.-G.\ de Gennes, J.\ of Stat.\ Phys.\ {\bf 119}, 953 (2005).

\bibitem{hummer}
A.\ Kaira, S.\ Garde, and G.\ Hummer, Proc.\ Nat.\ Acad.\ Sci.\ USA, {\bf 100} 10175 (2003). 


\bibitem{noy2}
F.\ Fornasiero {\em et al.}, Proc.\ Nat.\ Acad.\ Sci.\ USA, {\bf 105} 17250 (2008) .

\bibitem{corry}
B.\ Corry, J. Phys. Chem.\ B {\bf 112}, 1427 (2008).
  
\bibitem{schuster}
K. D. Kreuer, S. J. Paddison, E. Spohr and M. Schuster, Chem.\ Rev.\ {\bf 104}, 4637 (2004)

\bibitem{hummer2}
C. Dellago, M. M. Naor and G. Hummer, Phys.\ Rev.\ Lett.\ {\bf 90}, 10590 (2003).

\bibitem{hummer3}
C.\ Dellago and G.\ Hummer, Phys.\ Rev.\ Lett.\ {\bf 97}, 245901 (2006).

\bibitem{dellago1}
J.\ Köfinger, G.\ Hummer and C.\ Dellago, Proc.\ Natl Acad.\ Sci.\ USA {\bf 105}, 13218 (2008).

\bibitem{quantum}
N.\ Kavokine, M.-L.\ Bocquet, and L.\ Bocquet, Nature {\bf 602}, 84 (2022).

\bibitem{cui}
Cui et al., Sci.\ Adv.\ {\bf 11}, 7410 (2025).

\bibitem{Noy-Nature-Mat} 
Yuhao Li {\em et al.}, Nature Materials {\bf 23}, 1123–1130 (2024).

\bibitem{thole}
B.T.\ Thole, Chem.\ Phys. {\bf 59}, 341 (1981).

\bibitem{ashcroft} N. W. Ashcroft and N. D. Mermin, "Solid State Physics" (Holt, Rinehart and Winston, New York, 1976)

\bibitem{TF-sim} A. Schlaich, D. Jin, L. Bocquet and B. Coasne, Nature Materials 21,237 (2022).

%\bibitem{landau} 
%L.D.\ Landau and E.M.\ Lifshits, {\em Electrodynamics of a Contiuous Media} (Pergamon Press, New York, 1984). 

\bibitem{Ls} 
A.W.C.\ Lau and J.B.\ Sokoloff, Phys.\ Rev.\ E {\bf 111}, 045103 (2025).

\bibitem{yang} 
D. C. Yang,  et al, Nano Lett.\ {\bf 23}, 4956–4964 (2023).

%\bibitem{aluru} 
%Mohammad Heiranian , Amir Taqieddin , and Narayana R. Aluru, Phys.\ Rev.\ Res.\ {\bf 2}, 043153 (2020).

\bibitem{brothers} 
E.N.\ Brothers, E.N.\ Scuseria, G.E.\  Kudin,  J.\ Phys.\ Chem.\ B {\bf 110}, 12860–12864 (2006).

%\bibitem{ashcroft} 
N%.W.\ Ashcroft and N.D.\ Mermin, {\em Solid State Physics}  (Saunders Collage Publishing, 1976). 

\bibitem{charlier} 
J.-C. Charlier, X. Blase and S. Rosh, Reviews of Modern Physics 79, 677 (2007).


\bibitem{persson} 
B.N.J.\ Persson, J.\ Chem.\ Phys.\ {\bf 98}, 1695 (1993); Phys.\ Rev.\ B {\bf 44}, 3277 (1991); Surf. Sci. {\bf 103}, 269-270 (1992).

\bibitem{tobin} 
R.G.\ Tobin, Phys. Rev. B 48, 15468 (1993).

\bibitem{persson1} 
B.N.J.\ Persson, Phys. Rev. B {\bf 48}, 15471 (1993).

\bibitem{sokoloff} 
J.B.\ Sokoloff, Phys. Rev. B {\bf 52}, 5318 (1995).

\bibitem{boyer}
T.H.\ Boyer, Phys. Rev. A {\bf 9}, 68 (1974).

\bibitem{tomassone} 
M.S. Tomassone and A.\ Widom, Phys. Rev. B {\bf 56}, 4938 (1997); Am.\ J.\ Phys.\ {\bf 65}, 1181 (1997).

\bibitem{bruch} 
L.W.\ Bruch, Phys.\ Rev.\ B {\bf 61}, 16201 (2000); J.B.\ Sokoloff, J.\ Phys.: Condens.\ Matter {\bf 14}, 5277 (2002).

\bibitem{our-PRL} 
A.W.C.\ Lau and J.B.\ Sokoloff, Phys.\ Rev.\ Lett.\ 
{\bf 132}, 194001 (2024).

\bibitem{combre} 
Xuedan Ma, Sofie Cambré, Wim Wenseleers, Stephen K. Doorn, and Han Htoon, Phys.\ Rev.\ Let.\ {\bf 118}, 027402 (2017).

\bibitem{Tunuguntla} 
Ramya H. Tunuguntla, Frances I. Allen, Kyunghoon Kim1, Allison Belliveau and Aleksandr Noy, Nature Nanotechnology {\bf 11}, 639 (2016).

\bibitem{grotthus} 
C.J.T.\ Grotthus, Ann.\ Chim.\ LVIII, 54–74 (1806). 

\bibitem{dellago3} 
Jurgen Kofinger, Gerhard Hummera and Christoph Dellago, Phys. Chem.\ Chem.\ Phys.\ {\bf 13}, 15403 (2011). 


%%%%%%%%%%%%% above is last reference quoted %%%%%%%%%%%%%%%%

%\bibitem{dellago2}
%J. Köfinger and C. Dellago, Phys.\ Rev.\ Lett.\ {\bf 103}, 080601 (2009).

%\bibitem{roux}
%G. Lamoureux and B. Roux, J Chem.\ Phys., {\bf 119}, 3025-3039 (2003).


%\bibitem{blankschtein}
%R. P. Misra and D. Blankschtein, J Phys.\ Chem.\ C, {\bf 121} 28166 (2017).

%\bibitem{blankschtein2}
%R. P. Misra and D. Blankschtein, Langmuir, {\bf 37} 722-733 (2021).

%\bibitem{Toucette}
%A. Baule {\em et al.}, Nonlinearity, {\bf 24} 351 (2011).

%\bibitem{Toucette2}
%H.\ Touchette {\em et al.}, J. Phys. A:Math. Theor. {\bf 45}, 395002 (2012).

%\bibitem{Menzel}
%A.M.\ Menzel and N. Goldenfeld, Phys.\ Rev.\ E\ {\bf 84}, 011122 (2011).

%\bibitem{Fok-book}
%H.\ Risken, {\em The Fokker-Planck Equation.} (Springer-Verlag, Berlin, 1989).

%\bibitem{Reichl}
%L. E. Reichl, “A Modern Course in Statistical Physics,” (John Wiley and Sons, New York,
%1989), chapter 5.

%\bibitem{tom}
%P.M.\ Chaikin and T.C. Lubensky, {\em Principles of Condensed Matter Physics} (Cambridge University Press, New York, 1995).

%\bibitem{volokitin}
%A. I. Volokitin, Archiv, Cond. Mat. 2112.13752.







\end{thebibliography}
\end{document}